\documentclass[12pt]{article}
\usepackage{graphicx,psfrag,epsfig}
\usepackage{amssymb,amsmath,amscd,amsthm,mathtools}
\usepackage{graphicx,psfrag,epsfig}
\usepackage[english]{babel}
\usepackage{listings}

\usepackage{graphicx}
\usepackage{pstricks}
\usepackage{psfrag}

\usepackage[active]{srcltx}
\usepackage{setspace}  

\usepackage{url}
\usepackage{hyperref}

\DeclareMathOperator{\res}{Res}
\DeclareMathOperator{\diag}{Diag}
\DeclareMathOperator{\ddiag}{diag}
\DeclareMathOperator{\Tr}{Tr}
\DeclareMathOperator{\sgn}{sgn}
\DeclareMathOperator{\NCM}{NCM}
\DeclareMathOperator{\Sp}{Sp}
\DeclareMathOperator{\row}{row}
\DeclareMathOperator{\col}{col}

\date{}

\begin{document}

\date{}
\title{Adjoint Differentiation for generic matrix functions}
\author{ Andrei Goloubentsev\thanks{Barclays, NY, USA and MatLogica, London, UK., andrei.goloubentsev@matlogica.com}, Dmitri Goloubentsev\thanks{MatLogica, London, UK., dmitri@matlogica.com}, Evgeny Lakshtanov\thanks{CIDMA, Department of Mathematics, University of Aveiro, Portugal and MatLogica, London, UK., lakshtanov@matlogica.com}
}
\maketitle
\begin{abstract}
We derive a formula for the adjoint $\overline{A}$ of a square-matrix operation of the form $C=f(A)$, where $f$ is holomorphic in the neighborhood of each eigenvalue. We then apply the formula to derive closed-form expressions in particular cases of interest such as the case when we have a spectral decomposition $A=UDU^{-1}$, the spectrum cut-off $C=A_+$ and the Nearest Correlation Matrix routine. Finally, we explain how to simplify the computation of adjoints for regularized linear regression coefficients.
\end{abstract}

\tableofcontents

\section{Introduction}

This paper gives a general approach to calculating adjoints of functions $f(A)$ of square matrices. We complement the works of M.Giles \cite{giles} and B.N.Huge \cite{huge}, where they provide a list of closed-form formulas for adjoints of some useful matrix operations.

It is important to emphasize that we do not require the eigenvalues $\lambda_i \in \Sp(A)$ to be distinct and can apply the approach even when $A$ does not admit a spectral decomposition $A=UDU^{-1}$ by using more general normal form decompositions (e.g.  Schur or Jordan). However, we do require the decomposition to keep the identity matrix invariant. For instance, if $A$ is not symmetric, we can't use the singular value decomposition (SVD) \\ $A=UDV^T$ since $UV^T \neq Id$.


By the Cauchy's formulas we have

\begin{equation}\label{H}
f(z)=-\frac{1}{2\pi i} \int_{\Gamma} \frac{f(\lambda)}{z-\lambda} \,d \lambda, \quad f'(z)=\frac{1}{2\pi i} \int_{\Gamma} \frac{f(\lambda)}{(z-\lambda)^2} \,d\lambda,
\end{equation}
where $f$ is holomorphic on an open set containing $z$ and $\Gamma$ is the boundary of that set. These formulas will be crucial for calculating elements of the adjoint.

To derive the general formula for $\overline{A}$ we shall use the following generalization: 

\begin{equation} \label{HH}
f(A)=-\frac{1}{2\pi i} \int_{\Gamma} f(\lambda) \res(\lambda) \,d\lambda, \quad \res(\lambda)=(A-\lambda I)^{-1},
\end{equation}
where $f$ is holomorphic on an open set containing $\Sp(A)$, and $\Gamma$ is the boundary of that set. Note that we don't require the set to be connected so a union of disjoint discs around $\lambda_i \in \Sp(A)$ works. We refer to \cite{res}, \cite{dunford} for the proof\footnote{Note that a square matrix $A$ defines a linear operator $A: V \to V$ on a finite dimensional space $V$. The proof in the references works for a general class of linear operators, inluding the finite dimensional case.}.

{\bf Notation.} matrices in what follows are real-valued of size $n \times n$ unless stated otherwise. We use the standard notations $X^{-T}=(X^{-1})^{T}=(X^{T})^{-1}$ for the transpose of the inverse matrix of $X$, $X_{ij}$ for the $ij$ element of $X$ and $E_{ij}$ for the matrix with $1$ in the $ij$ position and zeroes elsewhere. We denote by $X.\row(i)$ the $i$-th row of $X$ and by $X.\col(j)$ the $j$-th column of $X$. We also denote by $A \circ B$ the component-wise (Hadamard) product $(A \circ B)_{ij}=A_{ij}B_{ij}$ of square matrices $A$ and $B$ of the same size. 

\section{The general formula for $\overline{A}$}

We first calculate the partial derivative $f'_{ij}(A)$ of $f(A)$ with respect to
$A_{ij}$. Note that for $\lambda \not \in \Sp(A)$ and $\varepsilon \to 0$ we have
$$
\left(A-\lambda +\varepsilon E_{ij}\right)^{-1}= \left[(A-\lambda)(I+\res(\lambda) \varepsilon E_{ij})\right]^{-1}=[I-\res(\lambda)\varepsilon E_{ij}+O(\varepsilon^2)] \res(\lambda)$$
and so
\begin{equation*}\label{D}
\res(\lambda)'_{ij}=- \res(\lambda) E_{ij} \res(\lambda).
\end{equation*}
Differentiating (\ref{HH}) with respect to $A_{ij}$ gives
$$
f_{ij}'(A) = -\frac{1}{2\pi i} \int_{\Gamma} f(\lambda) \res(\lambda)'_{ij} \,d\lambda = \frac{1}{2\pi i} \int_{\Gamma} f(\lambda) \res(\lambda) E_{ij} \res(\lambda) \,d\lambda,
$$
where we can take $\Gamma$ to be a set of small circles around each $\lambda_i \in \Sp(A)$ (since $f$ is holomorphic in the neighborhood each eigenvalue).

Following \cite{giles},  we can now calculate elements $\overline{A}_{ij}$ of the adjoint $\overline{A}$:
\begin{multline*}
\overline{A}_{ij} =\sum_{k,l} [f'_{ij}(A)]_{k,l} [\overline{f(A)}]_{k,l} = \Tr \left ( f'_{ij}(A) \overline{f(A)}^T \right ) = \\
 \frac{1}{2\pi i} \int_{\Gamma} f(\lambda) \Tr\left(\res(\lambda) E_{ij} \res(\lambda)\overline{f(A)}^T \right)d\lambda = \\
\frac{1}{2\pi i} \int_{\Gamma} f(\lambda) \Tr\left( E_{ij} \res(\lambda)\overline{f(A)}^T \res(\lambda)\right)d\lambda .
\end{multline*}
Since for any matrix $M$ one has $\Tr(E_{ij}M) =M_{ji}$, we get a general formula for $\overline{A}$:
\begin{equation}\label{A}
\overline{A}= \frac{1}{2\pi i} \int_{\Gamma} f(\lambda)   \res^T(\lambda)\overline{f(A)} \res^T(\lambda)d\lambda
\end{equation}

Now to obtain a closed-form expression for $\overline A$ it remains to transform $\res(\lambda)$ to an appropriate (keeping the identity matrix invariant) normal form and collect residues. We shall perform this computation in some cases of interest.

\section{Special cases}

\subsection{Spectral decomposition $A=UDU^{-1}$:}
Suppose that our matrix has a spectral decomposition $A=UDU^{-1}$ (including the case when not all the eigenvalues are distinct). Then $\res(\lambda) = U(D-\lambda I)^{-1}U^{-1}$ gives a (spectral) decomposition of $\res(\lambda)$. 

It is easy to see from (\ref{H}) that: 
\begin{equation} \label{L}
\frac{1}{2\pi i} \int_{\Gamma} \frac{f(\lambda)d\lambda}{(\lambda - \lambda_i)(\lambda - \lambda_j)} =
\begin{cases} \begin{aligned} &f'(\lambda_i), \quad &\lambda_i=\lambda_j \\
&\frac{f(\lambda_i)-f(\lambda_j)}{\lambda_i-\lambda_j} , \quad &\lambda_i \neq \lambda_j \end{aligned}
\end{cases} 
\end{equation}
Let $F$ be a matrix with these values: \begin{equation}\label{MM}
F_{ij} = \begin{cases} \begin{aligned} &f'(\lambda_i), \quad &\lambda_i=\lambda_j \\
&\frac{f(\lambda_i)-f(\lambda_j)}{\lambda_i-\lambda_j} , \quad &\lambda_i \neq \lambda_j
\end{aligned} \end{cases}
\end{equation}
Substituting $\res(\lambda) = U(D-\lambda I)^{-1}U^{-1}$ into (\ref{A}) and applying (\ref{L}) we see that \begin{equation*}\label{B}
\overline A^T= U \left(F \circ  (U^{-1} \overline{f(A)^T}  U)\right) U^{-1}.
\end{equation*}

We want to stress here that  $\overline{A}$ is expressed directly in terms of $\overline{f(A)}$ and there is no need to calculate adjoints to matrix decomposition components $U$ and $D$.

\subsection{Spectrum cut-off}\label{MMM}
Suppose now that $A$ is symmetric and $f(A)=A_+$ is the projector on the positive spectrum. Then we may integrate over a contour $\Gamma^+$ enclosing only the positive eigenvalues instead of integrating over $\Gamma$ (this follows from the Riesz decomposition theorem for operators, see e.g. \cite{hol},  \cite{gohberg}). Using the decomposition $A=UDU^T$ we immediately get 
\begin{equation}\label{C}
\overline{A}= U \left(F \circ  (U^{T} \overline{A_+}   U)\right) U^{T}
\end{equation}
with
\begin{equation} \label{G}
F_{ij}=\begin{cases}\begin{aligned} &\sgn(\lambda_i), \quad &\lambda_i=\lambda_j \\ 
&\frac{\max(\lambda_i,0)-\max(\lambda_j,0)}{\lambda_i-\lambda_j} , \quad &\lambda_i \neq \lambda_j 
\end{aligned} \end{cases}
\end{equation}
Note that (\ref{C}) can be generalized to other $f(A)$, as we will see later.

For smoothing purposes it is reasonable to use a smoothed indicator function $f(A)=\frac{1}{2}\left(I+\tanh\left(\frac{A}{\delta}\right)\right)$ instead of $f(A)=A_+$. In that case one should use (\ref{MM}) to define $F$.

\section{Nearest Correlation Matrix}\label{ncm}

\subsection{Setup}
Let $A \in \mathbb R^{n\times n}$ be a fixed symmetric matrix. Recall that the Nearest Correlation Matrix (NCM) is the solution $X$ to the following minimization problem: $$
\min \left \{\|A-X\|_F, ~ : ~ X=X^T, ~ X \geq 0, ~ \diag(X)=e \right \}.
$$
Here $\|\cdot\|_F$ is the Frobenius norm, $\diag(X)$ is the vector of diagonal elements of $X$, $e=\diag(Id)$ is the vector of $1$-s. By $X \geq 0$ we mean that any $\lambda \in \Sp(X)$ satisfies $\lambda \geq 0$ (the matrix is positive semidefinite).

Let $C=\NCM(A)$ be the Nearest Correlation Matrix. Then it is known that
\begin{equation*}\label{E}
C=\left(A+\ddiag(y_{*})\right)_+,
\end{equation*}
where $y_*=y_*(A)$  is a solution to the minimization problem
\begin{equation}\label{F}
\min_{y \in \mathbb R^n} \left(\frac{1}{2} \| (A+\ddiag(y))_+\|^2_F - e^Ty \right).
\end{equation}
Here, $\ddiag(y) , ~ y \in \mathbb R^n$ is the diagonal matrix with values on the diagonal coming from $y$ (so that we have $\diag(\ddiag(y))=y$). See \cite{qisun} for the proof of this result.

\subsection{Calculation}

Using results from section \ref{MMM} we will show  that 
$$
\begin{aligned}
\overline{A} &= \overline{A}_1+\overline{A}_2, \quad &\mbox{ where   } \\
\quad  \overline{A}_2 &= V\left(F \circ (V^T \overline{C} V)\right) V^T &\mbox{ and  } \\
\quad \overline{A}_1 &= -V\left(F \circ \left(V^T \ddiag\left(J^{-T}\diag(\overline{A_2}) \right) V\right)\right) V^T.
\end{aligned}
$$

Here $V$ and $F$ are defined similarly to $U$ and $F$ respectively in (\ref{C}), (\ref{G}) but for the operation $(A+\ddiag(y_*))_+$. These matrices are usually known at the forward pass, namely, at the last iteration of the Newton method for the minimization problem (\ref{F}).

The matrix $J=\{J_{ij}\}$ is given by\footnote{Another representation can be found in \cite{qisun}.}:
$$
J_{ij} = V\!.\!\row(i)  T^{ij}V^T\!\!.\! \col(j),$$
where $T^{ij}=\{[T^{ij}]_{kl}\}$ has elements $$[T^{ij}]_{kl}=V_{ik}V_{il} F_{kl}.
$$

{\bf Proof.}  The $\rm{NCM}$ routine consists of finding the solution  $y_*=y_*(A)$ of (\ref{F}) 
that also solves  $M(y)=e$, where  $$
M(y)=\rm{Diag}(A+\rm{diag}(y))_+.$$
Then $C=\rm{NCM}(A) = (A+\rm{diag}(y_*))_+$. So for the adjoint we have:
$$\begin{aligned}
\overline{A} &= \overline{A}_1+\overline{A}_2, \mbox{ where   } \\
\overline{A}_2 &=  V\left(F \circ (V^T \overline{C} V)\right) V^T
\end{aligned}
$$

Now we have to backpropagate through the solver which is straightforward  if we set the correct adjoint values before it starts (see e.g. \cite[(19),(20)]{uwe}). Namely, suppose that we have the solver backpropogate an implicit function $x=x(c)$ by equation $M(x,c)=0$. Then we should set $\overline{M}=-J^{-T}\overline{x}$ (where $J$ is the Jacobian) before backpropagating.

In our situation, a straightforward calculation using formula (\ref{C}) gives the $J$ above and setting 
$\overline{M}=-J^{-T}\rm{Diag}(\overline{A}_2) $ we get
$$
\overline{A}_1 = V\left(F \circ \left(V^T \ddiag(\overline{M}) V\right)\right) V^T.
$$
\qed

\section{Regression Regularization}\label{Regression}

\subsection{Setup}
Given a matrix $B$ of dimension $m \times k$ and  a matrix $A$ of  dimension $n \times m$, the standard formula for regression coefficients is
$$
\beta = (AA^T)^{-1}AB.
$$
In \cite{huge} B.N.Huge suggests computing regression coefficients $\beta$ using  spectrum cut-off with a spectrum threshold $\varepsilon \geq 0$ and a Tikhonov regularization parameter $\lambda \geq 0$:
\begin{equation}\label{CCC}
\beta = (AA^T)_{\varepsilon,\lambda}^{-1}AB,
\end{equation}
 where for  a positive semidefinite symmetric matrix $M=UDU^T$ we define $M_{\varepsilon,\lambda}^{-1}$ by 
$$
M_{\varepsilon,\lambda}^{-1} = U \ddiag  \left ( \frac{\sgn(\lambda_i-\varepsilon)}{\lambda_i+\lambda} \right ) U^T.
$$
Later in \cite{savin} this approach has been used by B.N.Huge and A.Savine to calculate stable adjoints for Callable Exotics prices.

\subsection{Calculation}

We only compute the adjoints for $M_{\varepsilon,\lambda}^{-1}$ since other operations in (\ref{CCC}) are matrix multiplications and corresponding formulas for adjoints  are simple and well-known (see \cite{giles}, \cite{huge}).
 
In \cite{huge} the author obtains adjoints of (\ref{CCC}) by calculating adjoints for elements of the SVD decomposition of the matrix $A$. We trust that our approach is easier to verify and the final expression is shorter than the formula \cite[(6)]{huge}. 

Indeed, (\ref{C}) generalizes to give the required expression for adjoints
\begin{equation}\label{CC}
\overline{M} = U \left(F \circ  (U^{T} \overline{M_{\varepsilon,\lambda}^{-1}}  U)\right) U^{T},
\end{equation}
where
$$
F_{ij} =  
\begin{cases} \begin{aligned} &\frac{-\sgn(\lambda_i-\varepsilon)}{(\lambda_i + \lambda)^2}, \quad &\lambda_i=\lambda_j \\ 
&\left ( \frac{\sgn(\lambda_i-\varepsilon)}{\lambda_i + \lambda} - \frac{\sgn(\lambda_j-\varepsilon)}{\lambda_j + \lambda} \right ) \frac{1}{\lambda_i-\lambda_j} , \quad &\lambda_i \neq \lambda_j \end{aligned}
\end{cases}
$$

 {\bf Acknowledgments.} We are thankful to Evgeny Ryskin and Antoine Savine for discussions on the subject. We would also like to thank Evgeny Goncharov for helping to prepare the final manuscript. E.Lakshtanov was partially supported by the Center for Research and Development in Mathematics and Applications (CIDMA) and the Portuguese Foundation for Science and Technology (``FCT--Funda\c{c}\~ao para a Ci\^encia e a Tecnologia'') within projects UIDB/04106/2020 and UIDP/04106/2020.

\end{document}